\documentclass[12pt]{article} 

\usepackage{cite,amssymb,latexsym}

\newcounter{thMM}
\newcounter{leMM}
\newcounter{deFF}
\newcounter{exMP}
\newenvironment{theorem}[1]{\refstepcounter{thMM}\trivlist
   \item[\hskip19pt{\sc #1~\arabic{thMM}.}]\it\hskip3pt}{\endtrivlist}
\newenvironment{lemma}[1]{\refstepcounter{leMM}\trivlist
   \item[\hskip19pt{\sc #1~\arabic{leMM}.}]\it\hskip3pt}{\endtrivlist}
\newenvironment{definition}[1]{\refstepcounter{deFF}\trivlist
   \item[\hskip19pt{\sc #1~\arabic{deFF}.}]\rm\hskip3pt}{\endtrivlist}

\title{\normalsize\bf GEOMETRIC (PRE)QUANTIZATION IN THE POLYSYMPLECTIC 
APPROACH TO FIELD THEORY\thanks{
To appear in ``Differential Geometry and its Applications'' 
(Proc. Conf., Opava (Czech Republic), August 27-31, 2001)  
Silesian University, Opava 2002.}  
} 

\author{
\normalsize 
 Igor V. Kanatchikov
\\
\normalsize 
II. Institut f\"ur Theoretische Physik 
der Universit\"at Hamburg \\
\normalsize      
Luruper Chaussee 149, 22761 Hamburg, Germany 
\\ 
\normalsize 
ivar@mail.desy.de\\[2ex] 
}

\date{\vspace*{-35pt}}

\catcode `\@=11
\@addtoreset{equation}{section}

\def\section{\@startsection {section}{1}{\z@}{-3.5ex plus -1ex minus
     -.2ex}{2.3ex plus .2ex}{\normalsize\bf}}
\def\subsection{\@startsection{subsection}{2}{\z@}{-3.25ex plus -1ex minus
 -.2ex}{1.5ex plus .2ex}{\normalsize\bf}}

\def\thebibliography#1{\section*{References\markboth
  {REFERENCES}{REFERENCES}}\list
  {[\arabic{enumi}]}{\settowidth\labelwidth{[#1]}\leftmargin\labelwidth
  \advance\leftmargin\labelsep
  \usecounter{enumi}}
  \def\newblock{\hskip .11em plus .33em minus -.07em}
  \sloppy
  \sfcode`\.=1000\relax}
 
\catcode `\@=12

\begin{document}

\maketitle

\begin{abstract}
The prequantization map for a Poisson-Gerstenhaber algebra 
of differential form 
  valued 
dynamical variables in the polysymplectic 
formulation of the De Donder--Weyl 
 covariant 
Hamiltonian field theory is presented and the 
corresponding prequantum Schr\"odin\-ger equation is derived.  
This is the first step toward understanding the procedures of 
precanonical field quantization 
from the point of view of geometric quantization. 
\medskip 


\noindent 
{\bf\small MS classification. } {\small 53D50, 81S10, 81T70, 70S05.} 

\end{abstract}


\newcommand{\beq}{\begin{equation}}
\newcommand{\eeq}{\end{equation}}
\newcommand{\beqa}{\begin{eqnarray}}
\newcommand{\eeqa}{\end{eqnarray}}
\newcommand{\nn}{\nonumber}

\newcommand{\half}{\frac{1}{2}}

\newcommand{\xt}{\tilde{X}}

\newcommand{\uind}[2]{^{#1_1 \, ... \, #1_{#2}} }
\newcommand{\lind}[2]{_{#1_1 \, ... \, #1_{#2}} }
\newcommand{\com}[2]{[#1,#2]_{-}} 
\newcommand{\acom}[2]{[#1,#2]_{+}} 
\newcommand{\compm}[2]{[#1,#2]_{\pm}}

\newcommand{\lie}[1]{\pounds_{#1}}
\newcommand{\co}{\circ}
\newcommand{\sgn}[1]{(-1)^{#1}}
\newcommand{\lbr}[2]{ [ \hspace*{-1.5pt} [ #1 , #2 ] \hspace*{-1.5pt} ] }
\newcommand{\lbrpm}[2]{ [ \hspace*{-1.5pt} [ #1 , #2 ] \hspace*{-1.5pt}
 ]_{\pm} }
\newcommand{\lbrp}[2]{ [ \hspace*{-1.5pt} [ #1 , #2 ] \hspace*{-1.5pt} ]_+ }
\newcommand{\lbrm}[2]{ [ \hspace*{-1.5pt} [ #1 , #2 ] \hspace*{-1.5pt} ]_- }

\newcommand{\pbr}[2]{ \{ \hspace*{-2.2pt} [ #1 , #2\hspace*{1.5 pt} ] 
\hspace*{-2.3pt} \} }
\newcommand{\nbr}[2]{ [ \hspace*{-1.5pt} [ #1 , #2 \hspace*{0.0pt} ] 
\hspace*{-1.5pt} ] }

\newcommand{\we}{\wedge}
\newcommand{\dv}{d^V}
\newcommand{\nbrpq}[2]{\nbr{\xxi{#1}{1}}{\xxi{#2}{2}}}
\newcommand{\lieni}[2]{$\pounds$${}_{\stackrel{#1}{X}_{#2}}$  }

\newcommand{\rbox}[2]{\raisebox{#1}{#2}}
\newcommand{\xx}[1]{\raisebox{1pt}{$\stackrel{#1}{X}$}}
\newcommand{\xxi}[2]{\raisebox{1pt}{$\stackrel{#1}{X}$$_{#2}$}}
\newcommand{\ff}[1]{\raisebox{1pt}{$\stackrel{#1}{F}$}}
\newcommand{\dd}[1]{\raisebox{1pt}{$\stackrel{#1}{D}$}}
\newcommand{\der}{\partial}
\newcommand{\oo}{$\Omega$}
\newcommand{\Om}{\Omega}
\newcommand{\om}{\omega}
\newcommand{\eps}{\epsilon}
\newcommand{\si}{\sigma}
\newcommand{\Lm}{\bigwedge^*}

\newcommand{\inn}{\hspace*{2pt}\raisebox{-1pt}{\rule{6pt}{.5pt}\hspace*
{0pt}\rule{.5pt}{8pt}\hspace*{4pt}}}
\newcommand{\sro}{Schr\"{o}dinger\ }
\newcommand{\bm}{\boldmath}
\newcommand{\vol}{\omega}
               \newcommand{\dvol}[1]{\der_{#1}\inn \vol}

\newcommand{\bd}{\mbox{\bf d}}
\newcommand{\bder}{\mbox{\bm $\der$}}
\newcommand{\bI}{\mbox{\bm $I$}}

\newcommand{\be}{\beta} 
\newcommand{\ga}{\gamma} 
\newcommand{\de}{\delta} 
\newcommand{\Ga}{\Gamma} 
\newcommand{\gmu}{\gamma^\mu}
\newcommand{\gnu}{\gamma^\nu}
\newcommand{\ka}{\kappa}
\newcommand{\hka}{\hbar \kappa}
\newcommand{\al}{\alpha}
\newcommand{\lapl}{\bigtriangleup}
\newcommand{\psib}{\overline{\psi}}
\newcommand{\Psib}{\overline{\Psi}}
\newcommand{\derts}{\stackrel{\leftrightarrow}{\der}}
\newcommand{\what}[1]{\widehat{#1}}

\newcommand{\bx}{{\bf x}}
\newcommand{\bk}{{\bf k}}
\newcommand{\bq}{{\bf q}}

\newcommand{\omk}{\omega_{\bf k}} 
\newcommand{\lpl}{\ell}
\newcommand{\zb}{\overline{z}} 


\normalsize 
 
\section{Introduction}

The idea of quantization of fields based on a manifestly 
covariant version of the Hamiltonian formalism in field theory 
known 
in the calculus of variation of multiple integrals 
\cite{dedonder,dedecker} 
has been proposed 
for several times throughout the last century dating back to 
M. Born and H. Weyl \cite{bw}. 
The mathematical study of 
geometrical structures underlying the related aspects of the 
calculus of variations and classical field theory 
has  been undertaken recently by several groups of authors 
\cite{gimm,norris,deleon,roemer,paufler1,roman,sardan,helein} 
including Demeter Krupka's group \cite{krupka} 
in the Czech Republic. 
One of the central issues for the purposes 
of quantization of fields is a proper definition of Poisson brackets 
within the covariant Hamiltonian formalism in field theory. 
This has been accomplished  in our earlier papers
\cite{romp98,bial96,goslar96} 
which are based on the notion of the {\em polysymplectic} form as 
a field theoretic analogue of the symplectic form in mechanics 
and present a construction of Poisson brackets 
of   differential forms 
 leading to a Poisson-Gerstenhaber algebra 
structure generalizing   a Poisson algebra in mechanics. 
The corresponding {\em precanonical\/} quantization of 
field theories was developed heuristically in 
\cite{qs96,bial97,lodz98}, its relation to the standard 
quantum field theory was considered in \cite{pla2001}, 
and a possible  application to quantum gravity 
was discussed in \cite{ijtp2001}. 
In this paper we present 
elements of geometric prequantization 
in field theory  
based on the abovementioned Poisson-Gerstenhaber brackets 
and derive the  corresponding prequantum analogue 
of the Schr\"odinger equation. 
The main purpose of our consideration is to pave a way to a better 
understanding of the procedures of precanonical field quantization 
 from the point of view of the principles of geometric quantization 
\cite{sniat}. 


\section{Polysymplectic structure and the Poisson-Gerstenhaber brackets} 

Let us briefly describe the polysymplectic structure  
\cite{romp98,goslar96}
which underlies the 
De Donder--Weyl (DW) Hamiltonian form of the field equations 
\cite{dedonder} 
\beq
\der_\mu y^a (x) = {\der H}/{\der p^\mu_a},  
\quad 
\der_\mu p^\mu_a (x) = - {\der H}/{\der y^a} ,  
\eeq 
where $p^\mu_a := \frac{\der L}{\der y^a_\mu}$, 
called {\em polymomenta\/}, 
and $H:= y^a_\mu p^\mu_a - L = H(y^a, p^\mu_a, x^\mu)$, 
called the {\em DW Hamiltonian function\/},  
are determined by the first order Lagrangian density 
$L = L(y^a, y^a_\mu, x^\nu)$.  These equations are 
known to be equivalent to the Euler-Lagrange field equations if 
$L$ is regular in the sense that 
$$\det \left ( \left |\left|  \frac{\der^2 L}{\der y^a_\mu \der y^b_\nu}
\right |\right | \right ) \neq 0.$$ 

Let us view classical fields $y^a= y^a(x)$ as sections in the 
{\em covariant configuration bundle} $Y\rightarrow X$ 
over an oriented $n$-dimensional space-time manifold $X$ with the volume form 
$\omega$. The local coordinates in 
 $Y$ are $(y^a,x^\mu)$. 
Let $\bigwedge{}^p_q(Y)$ denotes the space of 
$p$-forms on $Y$ which are annihilated by $(q+1)$ arbitrary vertical 
vectors of $Y$.

The space $\bigwedge^n_1(Y)\rightarrow Y$,    
which generalizes the cotangent bundle,    
is a model of 
a 
{\em multisymplectic phase space\/} \cite{gimm} 
which possesses the canonical structure 
\beq 
\Theta_{MS} = p_a^\mu dy^a \we \omega_\mu + p\, \omega , 
\eeq
where 
$\omega_\mu := \der_\mu\inn\omega$ 
are the basis of $\bigwedge^{n-1} T^*X$.  
The section $p= - H(y^a,p^\mu_a,x^\nu)$ gives the 
multidimensional 
{\em Hamiltonian Poincar\'e-Cartan form\/}  $\Theta_{PC}$.  
 
For the purpose of introducing the Poisson brackets 
which reflect the dynamical structure of DW Hamiltonian 
equations (1) we need a structure which is independent 
of $p$ or a choice of $H$: 

\begin{definition}{Definition} \label{def1} 
The {\em extended polymomentum phase space\/} is the quotient bundle 
$Z$: $\bigwedge^{n}_1(Y) / \bigwedge^{n}_0(Y) \!\rightarrow \! Y.  
$ 
\end{definition} 
The local coordinates on $Z$ are $(y^a,p_a^\nu,x^\nu)$. 
A canonical structure on $Z$ 
 can be understood as 
an equivalence class 
of forms 
$\Theta := [p_a^\mu dy^a\we \omega_\mu \quad {\rm mod} \bigwedge{}^{n}_0(Y)]$. 

\begin{definition}{Definition} \label{def2}  
The {\em polysymplectic structure\/} on $Z$ is an equivalence class 
 %
of 
forms $\Omega$ given by 
\beq
\Omega := [d\Theta \quad {\rm mod} \;\mbox{$\bigwedge^{n+1}_1(Y)$}] 
= [- dy^a\we dp^\mu_a \we \omega_\mu \quad {\rm mod}  
\;\mbox{$\bigwedge^{n+1}_1(Y)$}] . 
\eeq 
\end{definition}

\newcommand{\texxx}{ 

\begin{definition}{Definition} \label{def1} 
The 
quotient 
bundle 
$Z$: $\bigwedge^{n}_1(Y) / \bigwedge^{n}_0(Y) \!\rightarrow \! Y 
$ 
is called the {\em extended polymomentum phase space\/}. 
\end{definition} 
The local coordinates on $Z$ are $(y^a,p_a^\nu,x^\nu) =: (z^v, x^\mu)$. 
 
\begin{definition}{Definition} \label{def2}  
The {\em polysymplectic structure\/} on $Z$ is an equivalence class 
of closed non-degenerate forms $\Omega\in \bigwedge^{n+1}_2(Z)$ 
modulo $\bigwedge^{n+1}_1(Z)$.  
\end{definition} 
A suitable local coordinate expression of the polysymplectic form 
$\Omega$ is given by 
\beq
\Omega = - dy^a\we dp^\mu_a \we \omega_\mu 
\eeq 
which is understood as a representative in the 
quotient 
$\bigwedge^{n+1}_2(Z)/\bigwedge^{n+1}_1(Z)$. 
  } 

The equivalence classes are introduced as an alternative to 
the explicit introduction of a non-canonical connection 
on the multisymplectic phase space in order to define 
the polysymplectic structure as a ``vertical part'' of the 
multisymplectic form $d\Theta_{MS}$\cite{paufler1}. 
The fundamental constructions, 
such as 
the Poisson bracket below,  are 
required to be independent of the choice of representatives in the 
equivalence classes, as they are expected to be independent of the 
choice of a connection.

\begin{definition}{Definition} \label{def3}
A multivector field of degree $p$, $\xx{p}\in \bigwedge^p TZ$,  
is called {\em vertical\/} 
if $\xx{p}\inn F  = 0$ for any form $ F \in \bigwedge^{*}_0(Z)$. 
\end{definition} 

The polysymplectic form establishes a map of 
horizontal forms   of degree $p$, 
$\ff{p}$$\in$$\bigwedge^p_0(Z)$, $p=0,1,..., (n-1)$,
to vertical multivector fields 
of degree $(n-p)$, $\xx{n-p}{}_F$,  called {\em Hamiltonian\/}: 
\beq
\xx{n-p}{}_F\inn \Omega = d \ff{p}.  
\eeq 
More precisely, 
horizontal forms forms are mapped to the {\em equivalence 
classes\/} of Hamiltonian multivector fields 
modulo the {\em characteristic\/}   
multivector fields $\xx{p}_0$: $\xx{p}_0\inn\Omega = 0$, $p=2,...,n$. 
The forms for which the map (4) exists are also called {\em Hamiltonian\/}. 
It is easy to see that the space of Hamiltonian forms is not stable 
with respect to the exterior product of forms. However, 

\begin{lemma}{Lemma} \label{lem1} 
The space of Hamiltonian forms is closed with respect to the 
graded commutative, associative {\em co-exterior\/} product 
 \beq
\ff{p}\bullet \ff{q} := *^{-1}(*\ff{p}\we *\ff{q}) .  
\eeq 
\end{lemma} 

{\sc Proof: } A straightforward proof  is to solve (4) to see that 
Hamiltonian $p$-forms are restricted to specific  $(n-p)$-polylinear 
forms in $p^\mu_a$, and then to check that the $\bullet-$product 
preserves the space of these forms (see \cite{bial96,paufler1}). 
{\hfill $\Box$}  

\medskip 

Note that  
 the definition of the  $\bullet$-product requires only 
the volume form $\omega$ on the space-time, not the metric structure. 
Given $\omega$ a $p$-form $F \!\in\! \bigwedge^p T^*X$ can be mapped to 
an $(n-p)-$multivector $X_F\!\in\! \bigwedge^{n-p} TX$: 
$X_F\inn \omega=F$. Then the 
exterior product of multivectors $(\wedge)$ induces   the $\bullet$-product 
of forms in $\bigwedge^* T^*X$. 
The construction 
is given by 
the 
commutative diagram 
$$
\begin{array}{ccc}
& \quad \bullet \quad & \vspace*{-6pt }\\
\mbox{$\bigwedge^p$} T^*X \otimes  \mbox{$\bigwedge^q$}T^*X& 
$\rightarrowfill$ & \mbox{$\bigwedge^{p+q-n}$} T^*X \vspace*{+5pt }
\\
\!\!\!\!\! \omega\ \Big\downarrow  & &\Big\downarrow \ \omega 
\\
& \quad \wedge \quad \vspace*{-5pt }&
\\
\mbox{$\bigwedge^{n-p}$}TX \otimes\mbox{$\bigwedge^{n-q}$}TX  
& $\rightarrowfill$ & \mbox{$\bigwedge^{n-p+n-q}$} TX
\end{array}
$$                                                            
and can be lifted to forms in $\bigwedge^*_0(Z)$.

The Poisson bracket of Hamiltonian forms
 $\pbr{\;}{\,}$  is induced by the 
Schouten-Nijenhuis bracket $\nbr{\;}{\,}$ 
of the corresponding 
Hamiltonian multivector fields: 
\beq
- d \pbr{\ff{p}}{\ff{q}} := 
\nbr{\xx{n-p}}{\xx{n-q}} \inn \Omega . 
\eeq
As a consequence, 
\beq
\pbr{\ff{p}{}_1}{\ff{q}{}_2} = (-1)^{(n-p)} \xx{n-p}{}_1 \inn d \ff{q}{}_2 
= (-1)^{(n-p)} \xx{n-p}{}_1 \inn \xx{n-q}{}_2 \inn \Omega , 
\eeq 
whence the independence of the definition of the choice of 
representatives  in the equivalence classes of $X_F$ and $\Omega$ 
is obvious. The algebraic properties of the bracket 
are given by the following 

\begin{theorem}{Theorem} \label{the1} 
The space of Hamiltonian forms with the 
operations $\pbr{\;}{\,}$ and $\bullet$  
is  a { (Poisson-)Gerstenhaber algebra\/}, i.e. 
\beqa
\pbr{\ff{p} }{\ff{q} } &=& -(-1)^{g_1 g_2}
\pbr{\ff{q}}{\ff{p}}, \nn \\ 
\mbox{$(-1)^{g_1 g_3} \pbr{\ff{p}}{\pbr{\ff{q}}{\ff{r}}}$} 
&\!+\!& 
\mbox{$(-1)^{g_1 g_2} \pbr{\ff{q}}{\pbr{\ff{r}}{\ff{p}}}$} 
  \\
&& \hspace*{15pt}+ \quad \! 
\mbox{$(-1)^{g_2 g_3} \pbr{\ff{r}}{\pbr{\ff{p}}{\ff{q}}} 
 \; = \;0, $}  
 \nn \\
\pbr{\ff{p}}{\ff{q}\bullet \ff{r}} 
&=& 
\pbr{\ff{p}}{\ff{q}}\bullet \ff{r}
+ (-1)^{g_1(g_2+1)} \ff{q}\bullet\pbr{\ff{p}}{\ff{r}},   
\nn 
\eeqa 
where $g_1 = n-p-1$,  $g_2 = n-q-1$,  $g_3 = n-r-1$.  
\end{theorem} 

{\sc Proof: } The graded Lie algebra properties are a 
straightforward consequence of (2.6) and the graded Lie nature of the 
Schouten-Nijenhuis bracket. The graded Leibniz property can be 
seen as a consequence of the Fr\"olicher-Nijenhuis theorem \cite{fn}
adapted to the algebra of forms equipped with the co-exterior 
product. 
{\hfill $\Box$}

\section{Prequantization map} 

Having in our disposal a generalization of the symplectic structure and 
a Poisson algebra to 
the DW Hamiltonian formalism  
of field theory        
it is natural to ask if geometric quantization 
can be generalized to this framework. The first step in this 
direction would be a generalization of the {\em prequantization map\/}  
 \cite{sniat}
$F \rightarrow O_F$ 
which maps dynamical variables $F$ on the classical phase 
space to the first order (prequantum) operators $O_F$ 
on (prequantum) Hilbert space and fulfills three properties: 
  \begin{quote}
(Q1) the map $F\rightarrow O_F$ is linear; \\
(Q2) if $F$ is constant, then $O_F$ is the corresponding multiplication 
operator; \\
(Q3) the Poisson bracket of dynamical variables is 
related to the commutator of the corresponding operators 
as follows: 
\end{quote} 
\beq 
[O_{F_1}, O_{F_2}]=-i\hbar O_{\{F_1, F_2\}}.   
\eeq  
In the case of a Poisson-Gerstenhaber algebra we expect that the commutator 
(1) is replaced by the {\em graded\/} commutator 
$[A,B] := A\co B - (-1)^{\deg A \deg B} B\co A.$ 

\begin{theorem}{Theorem} \label{TH3}
The prequantum operator of a differential form 
dynamical variable $F$ 
is given by the formula 
\beq
O_{F}= i\hbar \pounds_{X_F} 
+ X_F{} \inn \Theta \bullet + F\bullet ,  
\eeq 
where $\pounds_{X_F} := [X_F, d]$  and 
$\Theta$ is a 
(local) polysymplectic potential 
 in the sense of (2.3). 
 \end{theorem}

{\sc Proof: } See a straightforward calculation in 
\cite{torun2001}. {\hfill $\Box$}  \\
 
The most intriguing aspect of the representation (2) is that the 
prequantum operator $O_F$ is non-homogeneous: 
for an $f$-form $F$ the degree of the first term in (2) 
is  $(n-f-1)$ and the degree of the other two terms is $(n-f)$. 
This 
 fact 
suggests that  prequantum wave functions are 
 complex 
non-homogeneous horizontal differential forms, 
i.e. 
sections of the complexified bundle 
\mbox{$\bigwedge^*_0(Z)^{\mathbb C} \rightarrow Z$}. 
The corresponding (graded) prequantum Hilbert space 
 will be considered in \cite{torun2001} (see also \cite{mg9a}).  

Note that formulas (1) and (2) imply  that one can introduce a formal 
non-homogeneous ``supercovariant derivative'' with respect to a 
multivector field 
$X$: $\nabla_X:= \pounds_X -\frac{i}{\hbar} X\inn\Theta\bullet$ 
with the curvature of the corresponding ``superconnection'' $\nabla$ 
(cf. \cite{quillen})  
\beq
\Omega(X_1, X_2) := -i\hbar \left ( 
[\nabla_{X_1}, \nabla_{X_2}] - \nabla_{\nbr{X_1}{X_2}} \right ) 
\eeq 
coinciding with the polysymplectic form. 

One of the important questions is what is the dynamical equation 
for the wave functions. Let us consider how geometric 
prequantization can help us to find an answer.

\section{Prequantum Schr\"odinger equation} 
 
The origin of the Schr\"odinger equation in quantum mechanics 
from the point of view of geometric (pre)quantization can be understood 
as follows. The classical equations of motions are incorporated in the 
vector field $X_*$ which annihilates the exterior differential of the 
(Hamiltonian) Poincare-Cartan form 
\beq
\Theta = p dq - H(p,q) dt , 
\eeq 
i.e. 
\beq
X_*\inn d\Theta =0. 
\eeq
The classical trajectories in the phase space are known to be 
the integral curves of $X_*$. 

Now, if we think of geometric prequantization based on the presymplectic 
structure given by $d\Theta$ we notice 
that the zero ``observable'' has a non-trivial 
(presymplectic) prequantum operator: 
\beq 
 0 \rightarrow O_0 = i\hbar \pounds_{X_*} + X_*\inn \Theta , 
\eeq 
where 
$ 
X_*=X^t\der_t + X^q\der_q + X^p\der_p  
$ 
and 
\beq
X^q = \der_p H, \quad X^p = -\der_q H , 
\eeq 
as it follows from (2) under the assumption $X^t=1$ (which is just a 
choice of time parametrization). The obvious consistency requirement 
then is that $O_0$ vanishes on prequantum wave functions 
$\Psi=\Psi (p,q,t)$, i.e. 
\beq   
O_0 (\Psi) = 0 . 
\eeq 
Using the explicit form of the operator $O_0$ derived from (3), (4): 
\beq
O_0=  i\hbar \der_t + i\hbar (\der_p H \der_q - \der_q H \der_p) 
+ p\der_p H - H(p,q) 
\eeq 
one can write (5) in the form of the 
{\em prequantum Schr\"odinger equation\/}
\beq
i\hbar \der_t \Psi = O_H \Psi , 
\eeq 
where $O_H$ is the (symplectic) prequantum operator of 
the Hamilton canonical function:  
\beq
O_H = - i\hbar (\der_p H \der_q - \der_q H \der_p) 
 - p\der_p H + H(p,q)  . 
\eeq 

The above consideration demonstrates 
the origin of the Schr\"odinger equation in 
the classical relation (2) . The subsequent 
steps of quantization just reduce the Hilbert space of the wave 
functions (by choosing a {\em polarization}) 
and construct a proper operator of $H$ on this Hilbert space, 
the form of the Schr\"odinger equation (7) remaining intact.   
This observation motivates our consideration of the field 
theoretic prequantum Schr\"odinger equation in the following 
section: having obtained it on the  level of prequantization 
one may have a better idea as to what is the covariant 
Schr\"odinger equation for quantum fields within the approach 
based on DW Hamiltonian formulation (2.1). 

There has been a little discussion of the 
prequantum Schr\"o\-dinger equation in the literature  
(cf. \cite{preschr})  
for the reason that it works on a wrong 
Hilbert space of functions over the phase space, thus 
contradicting the uncertainty principle. 
It can serve, therefore, only as an intermediate step toward 
the true quantum mechanical Schr\"odinger equation. 

Let us note that 
eqs. (7), (8) recently 
appeared within the hypothetical 
framework  of ``subquantum mechanics'' proposed by J. Sou\v cek 
\cite{soucek} 
whose starting point was quite different from geometric quantization.  
A possible 
connection between the  ``subquantum mechanics''  and geometric 
prequantization could be an interesting subject to study, 
particularly in connection with the question recently 
revisited by G. Tuynman \cite{tuynman2} as to ``were 
there is the border between classical and quantum mechanics in geometric 
quantization?''

\section{Prequantum Schr\"odinger equation in field theory} 

In this section we present a field theoretic generalization of the 
above derivation of the prequantum Schr\"odinger equation. 

It is known \cite{roemer,roman,romp98}  
that the classical field equations 
 in the form (2.1) 
are encoded in the multivector field 
of degree $n$, $\xx{n}{}_* \in \bigwedge^n TZ$, 
which annihilates the exterior differential 
of the multidimensional Hamiltonian Poincare-Cartan $n$-form 
\beq
\Theta_{PC} = p^\mu_a dy^a \we \omega_\mu - H(y^a, p_a^\mu) \omega , 
\eeq
i.e. 
\beq
\xx{n}{}_* \inn d\Theta_{PC}= 0.  
\eeq


Let us extend the geometric prequantization map (3.2) to the 
case of the ``pre-polysymplectic'' form $d\Theta_{PC}$,  
 usually called {\em multisymplectic\/}. 
Again, the feature of this extension is that there is a 
non-trivial prequantum operator corresponding to the 
zero function on the polymomentum phase space:  
\beq 
O_0 = i\hbar [\xx{n}{}_*, d] + \xx{n}{}_* \inn \Theta_{PC}\bullet. 
\eeq 
Therefore, the consistency requires the prequantum wave 
function $\Psi = \Psi(y^a, p^\mu_a,x^\mu)$ to obey the condition 
\beq 
O_0 (\Psi) =0 
\eeq 
which is expected to yield the field theoretic 
prequantum Schr\"odinger equation.

It is easy to see that the operator (3) is non-homogeneous: the first 
term has the degree $-(n-1)$ while the last one has the degree $-n$. 
Therefore, the prequantum wave function in (4) is a 
 horizontal 
non-homogeneous form 
$$
\Psi = \psi \omega + \psi^\nu\omega_\nu,  
$$ 
a section of the bundle  
$ \left ( \bigwedge^{n-1}_0(Z) \oplus \bigwedge^{n}_0(Z) 
\right ){}^{\mathbb C}   \rightarrow Z$ which generalizes the complex line bundle 
over the symplectic phase space used in the usual geometric quantization.

In terms of the Hamiltonian vector field associated with $H$: 
$$  
\xx{n}{}_H \inn d\Theta = dH, 
$$ 
where $\Theta$ is a potential of the polysymplectic form $\Omega$,
the vertical part of $\xx{n}{}_*$ takes the form: 
$$  
\xx{n}{}_*^V= (-1)^n (\xx{n}{}_*\inn \omega) \xx{n}_H. 
$$
Then (5.4) yields the prequantum Schr\"odinger equation in 
the form: 
\beq
i\hbar (\der_\mu \psi^\mu - 
(-1)^n \der_\mu \psi \, dx^\mu)  
=  - (-1)^{n} \left ( i\hbar \xx{n}{}_H \inn d \Psi 
+ \xx{n}{}_H \inn \Theta  \bullet \Psi \right )    
+ H\bullet \Psi.  
\eeq
   
\newcommand{\oldtextl}{
Assuming $\xx{n}_*\inn \omega = const$ and introducing the 
vertical Hamiltonian multivector field associated with $H$: 
$$  
\xx{n}{}_H \inn d\Theta = dH, 
$$ 
where $\Theta$ is a potential of the polysymplectic form $\Omega$, 
the prequantum Schr\"odinger equation (4) takes the form: 
\beq
i\hbar (\der_\mu \psi^\mu - 
(-1)^n \der_\mu \psi \, dx^\mu)  
=  - (-1)^{n} i\hbar \xx{n}{}_H \inn d \Psi 
+ 
(-1)^{n} \xx{n}{}_H \inn \Theta  \bullet \Psi   
+ H\bullet \Psi.  
\eeq
 }

For {\em odd\/} $n$ the right hand side of (5) is identified with 
the  (polysymplectic) prequantum operator of $H$ (see (3.2)),   
and (5) takes a particularly appealing form (cf. \cite{bial94}) 
\beq 
i\sigma \hbar \, d\bullet \Psi = O_H (\Psi) , 
\eeq 
where $\sigma = \pm 1$ for Euclidian/Lore\-ntzian spacetimes 
(in our conventions $*\omega=\sigma$), 
and $d\bullet$ is the {\em co-exterior\/} differential \cite{bial97} 
which is non-vanishing only on the subspace of $(n-1)$- and $n$-forms: 
$$d\bullet (\psi^\nu \omega_\nu) = 
\der_\mu \psi^\nu dx^\mu\bullet \omega_\nu = \sigma \der_\nu \psi^\nu, 
\quad 
d\bullet (\psi \omega) = \sigma \der_\mu \psi dx^\mu.$$ 

For {\em even\/} $n$ the right hand side of (5) 
is not $O_H$ because of the wrong sign 
in front of the first two terms. The left hand side 
is also different from the one in (6). 
 A distinction between even and odd space-time dimensions is a problematic 
 feature of the present derivation 
based on a specific prequantization 
formula (3.2).\footnote{See Note added in proofs.} 

The meaning of our discussion in this section is that it provides 
a hint to the actual form of the covariant Schr\"odinger 
equation 
in field theory 
which one can expect within 
 the approach to 
field quantization 
based on 
the covariant DW Hamiltonian formalism.


\newcommand{\oldtextf}{
Assuming for simplicity that 
\beq
X\uind{\mu}{n} \der\lind{\mu}{n} \inn \omega = 1 
\eeq

we obtain: 
\beqa 
[\xx{n}{}^h_*, d] \Psi &=& \der_\mu \psi^\mu - 
(-1)^n \der_\mu \psi \, dx^\mu, 
\nn \\ 
{}[ \xx{n}{}^V_*, d ] \Psi &=& (-1)^{n-1} \xx{n}{}^v{}\uind{\mu}{n-1} 
\der\lind{\mu}{n-1} \inn (\der_v \psi \,\omega + \der_v \psi^\mu \,\omega_\mu) , 
\nn \\ 
{}[\xx{n}{}^{VV}_*, d] \Psi &=& 0,  
\quad \mbox{\rm  etc. } 
\nn 
\eeqa
Besides, 
$$
\xx{n}{}^{}_* \inn \Theta_{PC} = 
{}\xx{n}{}^{V}_* \inn \Theta_{PC} - H .  
$$ 
} 

\section{Discussion} 

We presented a formula of prequantum operators  
corresponding to Hamiltonian forms. It realizes a 
representation of the Poisson-Gerstenhaber algebra 
of Hamiltonian forms by operators acting 
on prequantum 
wave functions given by 
nonhomogeneous forms  $\Psi$, the sections of 
$\bigwedge^*_0(Z)^{\mathbb C}   \rightarrow Z$.  
We also argued that these wave functions  
fulfill the prequantum Schr\"odinger equation (5.5). 


The next step in geometric quantization would be to reduce the 
prequantum Hilbert space by introducing a {\em polarization\/} in the 
polymomentum phase space.   
A generalization of the {\em vertical\/} polarization 
reduces the space of wave functions 
to the functions depending on 
field variables and space-time variables: $\Psi (y^a, x^\mu)$. 
A construction of 
quantum operators on the new Hilbert space 
of quantum wave functions requires further generalization 
of the techniques of geometric quantization,  
such as the notion of the metaplectic correction and the 
Blattner-Kostant-Sternberg pairing,  
which is not developed yet. 

However, the quantum operator $\what{H}$ 
is already known from the heuristic 
procedure of ``precanonical quantization'' 
\cite{qs96,bial97,lodz98,pla2001,ijtp2001}  
based on  quantization of a small 
Heisenberg-like subalgebra of brackets of 
differential forms generalizing the canonical variables.  
Within  precanonical quantization it was found suitable 
to work in terms of the space-time Clifford algebra valued 
operators and wave functions, rather than in terms of 
non-homogeneous forms and the graded endomorphism valued 
operators acting on them. In general, 
a relation between the two formulations is given by the 
``Chevalley quantization'' map from the 
{\em co-\/}exterior algebra to the Clifford algebra: 
$\omega_\mu\bullet \rightarrow 
- \frac{1}{\varkappa} \gamma_\mu,$ where the constant $\kappa$ 
is introduced to match the physical dimensions 
($1/\varkappa \sim$ length{}$^{n-1}$). 
The  corresponding Clifford product of forms is  
given by (cf. \cite{joos}) 
$$\omega_\mu\vee\omega_\nu = 
\omega_\mu\bullet\omega_\nu + \varkappa^{-2} \eta_{\mu\nu}.
$$
Note that the appearence of the metric $\eta_{\mu\nu}$ 
at this stage is related to the fact that a definition 
of the scalar product of wave functions 
represented by non-homogeneous forms, i.e. their probabilistic 
interpretation, requires a space-time metric. 

Under the above ``Cliffordization'' and the vertical polarization 
the wave function becomes Clifford valued: $\Psi=\Psi(y^a,x^\mu)$ 
and the left hand side of  (5.5), (5.6) can be expressed in terms of the 
Dirac operator acting on $\Psi=\psi + \psi_\nu\gamma^\nu$;  
in particular, $d\bullet \sim \gamma\gamma^\nu\der_\nu$, 
where 
 $\gamma \sim \gamma_1\gamma_2...\gamma_n$ 
corresponds to the Hodge 
duality operator $*$.  
Similarly, 
the operator of $H\bullet$  is represented as $\sim \!\gamma \what{H}$. 
The coefficients not specified here are fixed by the requirement that 
the resulting Dirac-like equation is causal and consistent, 
thus leading to the covariant Schr\"odinger equation for 
quantum fields in the form 
\beq
i\hbar\varkappa \gamma^\mu\der_\mu \Psi = \what{H}\Psi . 
\eeq 
A similar reasoning leads to the represenation of polymomenta: 
$\hat{p}{}_a^\mu = -i\hbar\varkappa \gamma^\mu {\der}/{\der y^a}$.  
These results have been anticipated  within precanonical 
field quantization earlier \cite{qs96,bial97,lodz98} 
(see also \cite{navarro} where similar relations were postulated). 
This approach also allows us to derive the explicit form of 
$\what{H}$. For example,  
in the 
case of interacting scalar fields $y^a$ one can show that \cite{bial97} 
$$ 
\what{H} = - \mbox{\large $\frac{1}{2}$} \hbar^2\varkappa^2 
\lapl + V(y), 
$$ 
where $\lapl$ is the Laplace operator in the space of field 
variables.

\newcommand{\oldtextk}{
Under this ``Cliffordization'' 
the Schr\"odinger equation on $\Psi=\Psi(y^a,x^\mu)$ 
can be argued to take the  form  (cf. (5.6)): 
\beq
i\hbar\varkappa \gamma^\mu\der_\mu \Psi = \what{H}\Psi 
\eeq 
and the polymomenta are represented as follows: 
$\hat{p}{}_a^\mu = -i\hbar\varkappa \gamma^\mu {\der}/{\der y^a}.$  
In the particular case of interacting 
scalar fields $y^a$, one can show that \cite{bial97} 
$$ 
\what{H} = -\mbox{$ \frac{1}{2}$} \hbar^2\varkappa^2 
\lapl + V(y), 
$$ 
where $\lapl$ is the Laplace operator in the space of field 
variables. 
 } 

What we have arrived at is a multidimensional hypercomplex 
generalization of the Schr\"odinger equation from quantum mechanics 
to field theory, where the space-time Clifford algebra, which 
arose from quantization of differential forms, generalizes the 
algebra of the complex numbers in quantum mechanics, and the notion of the 
unitary time evolution is replaced by the 
space-time propagation  governed by the Dirac operator. 
In \cite{pla2001} we discussed how this description of quantum 
fields can be related to the standard description in the 
functional Schr\"odinger representation. In doing so the 
Schr\"odinger wave functional arises as a specific composition 
of amplitudes given by Clifford-valued wave functions of the 
precanonical approach, and the 
parameter $\varkappa$ appears to be related to the 
ultra-violet cutoff. 

Obviously, in this presentation we have left untouched a lot 
of important issues both on the level of prequantization 
and on the level of quantization.  A development of 
the present version of geometric quantization in field theory 
would  further clarify the mathematical foundations of 
precanonical quantization of fields and also advance its 
understanding and applications.   The whole field appears 
to us as appealing, mathematically rich and unexplored 
as the field of 
 the 
geometric quantization 
approach to quantum mechanics 
was 25-30 years ago. 

\bigskip

{ 

{\bf Note added in proofs (May, 2002)}   

\medskip 

A distinction between odd and even $n$ in Sect. 5 can be avoided 
by noticing that the prequantization map (3.2) can be modified 
as follows: 
\beq
O'_{F}= (-1)^{(n-f-1)}\left ( i\hbar \pounds_{X_F} 
+  X_F{} \inn \Theta \bullet \right ) 
+ F\bullet ,
\eeq
where $(n-f)$ is the co-exterior degree of $F$. 
Then the right hand side of 
(5.5) is identified with $O'_{H}(\Psi)$ for any $n$. The left 
hand side of (5.5) also can be written in a universal form for 
any $n$ using the 
reversion anti-automorphism $\beta$
in a co-exterior Grassmann algebra:   
$\beta(F) := (-1)^{\frac{1}{2}(n-f)(n-f-1)}F$. 
Then the prequantum Schr\"odinger equation (5.5) 
can be written as follows: 
\beq
i\sigma\hbar (-1)^{\frac{1}{2}n(n-1)}\beta(d\bullet\Psi) = O'_{H} (\Psi). 
\eeq
Note that a choice between two representations (3.2) and (6.2) 
can be made once the scalar product is specified. 

} 

\bigskip 

\medskip 

{\bf Acknowledgments. } 

\medskip 

The work has been supported by  
the 
{\em Graduiertenkolleg\/} 
``Future Developments in Particle Physics'' 
at the University of Hamburg 
to which I express my sincere gratitude.

\end{document}